\documentclass[conference]{IEEEtran}
\IEEEoverridecommandlockouts
\usepackage{cite}
\usepackage{amsmath,amssymb,amsfonts}
\usepackage{algorithmic}
\usepackage{graphicx}
\usepackage{textcomp}
\usepackage{xcolor}
\def\BibTeX{{\rm B\kern-.05em{\sc i\kern-.025em b}\kern-.08em
    T\kern-.1667em\lower.7ex\hbox{E}\kern-.125emX}}

\usepackage{booktabs}

\usepackage{comment}

\usepackage[inkscapearea=drawing]{svg}

\usepackage{bm}

\usepackage{url}

\usepackage{cite}

\usepackage{xcolor}

\usepackage{stackrel}

\usepackage{todonotes} 

\usepackage{lipsum}
\usepackage{algorithm}
\usepackage{siunitx}

\usepackage{epsfig} 
\usepackage[T1]{fontenc}

\usepackage{mathtools}
\usepackage{csquotes}
\usepackage{xcolor}  %

\usepackage{pgf}
\usepackage{pgfplots} %
\usepackage{adjustbox} %

\usepackage{bbm}
\usepackage{amsbsy}
\usepackage{amssymb}
\usepackage{amsmath}
\usepackage{exscale} %
\usepackage{icomma} %
\usepackage{mathrsfs} %
\usepackage{array} %
\usepackage{multirow} %
\usepackage{listings} %
\usepackage{relsize}
\usepackage[inline, shortlabels]{enumitem}
\usepackage{dsfont}

\newcommand\blfootnote[1]{%
  \begingroup
  \renewcommand\thefootnote{}\footnote{#1}%
  \addtocounter{footnote}{-1}%
  \endgroup
}

\usetikzlibrary{positioning}
\usetikzlibrary{positioning, shapes.geometric} 
\usetikzlibrary{external}
\tikzsetexternalprefix{tikz_fig/} 

\usepackage[caption=false,font=footnotesize]{subfig}
\usepackage{mathtools,amssymb,tikz,bm,siunitx, booktabs}
\usepackage[colorlinks=false,bookmarks=false,citecolor=black,urlcolor=blue]{hyperref} %

\definecolor{KITgreen}{rgb}{0,.59,.51}
\definecolor{KITorange}{rgb}{.87,.60,.10}
\definecolor{KITred}{rgb}{.63,.13,.13}
\definecolor{KITpurple}{RGB}{160,0,120}
\definecolor{KITcyanblue}{RGB}{80,170,230}

\newcommand{\argmax}{\text{arg}\,\text{max}}

\begin{document}

\title{Spiking Neural Network \\ Decision Feedback Equalization}

\author{\IEEEauthorblockN{Eike-Manuel Bansbach, Alexander von Bank, and Laurent Schmalen}
\IEEEauthorblockA{Communications Engineering Lab, Karlsruhe Institute of Technology, 76187 Karlsruhe, Germany \\
(email: \texttt{e.bansbach@kit.edu, alexander.bank@kit.edu})}
}

\maketitle

\begin{abstract}
In the past years, artificial neural networks (ANNs) have become the de-facto standard to solve tasks in communications engineering that are difficult to solve with traditional methods. 
In parallel, the artificial intelligence community drives its research to biology-inspired, brain-like spiking neural networks (SNNs), which promise extremely energy-efficient computing. 
In this paper, we investigate the use of SNNs in the context of channel equalization for ultra-low complexity receivers. 
We propose an SNN-based equalizer with a feedback structure akin to the decision feedback equalizer (DFE). 
To convert real-world data into spike signals, we introduce a novel ternary encoding and compare it with traditional log-scale encoding. 
We show that our approach clearly outperforms conventional linear equalizers for three different exemplary channels.
We highlight that mainly the conversion of the channel output to spikes introduces a minor performance penalty. 
The proposed SNN with a decision feedback structure enables the path to competitive energy-efficient transceivers.
\end{abstract}

\section{Introduction}
\blfootnote{This work has received funding from the European Research Council (ERC) under the European Union’s Horizon 2020 research and innovation programme (grant agreement No. 101001899).
Parts of this work were carried out in the framework of the CELTIC-NEXT project AI-NET-ANTILLAS (C2019/3-3) and were funded by the German Federal Ministry of Education and Research (BMBF) under grant agreement 16KIS1316.}%
The innovations in communications engineering in the past decades have led to transmitters and receivers that are extremely powerful and can provide near-capacity transmission in various transmission scenarios.
One reason is the recent use of machine learning techniques, more specifically artificial neural networks (ANNs), that can mitigate device and transmission impairments and solve specific tasks that are difficult to solve with traditional methods. In this paper, we focus on the problem of channel equalization, which can profit largely from the use of machine learning and ANNs, see, e.g.,~\cite{Carrera21,Pedro22,Lauinger22}.

In many scenarios, the ANN-based equalizer's performance depends on its computational complexity; hence their use in ultra-low-complexity systems may be prohibitive~\cite[Fig.~6]{Pedro21}.
However, with increasing computational complexity, the overall energy consumption of the system grows, leading to power-hungry receivers when implemented on digital electronics.
Instead of digital electronics, the use of so-called neuromorphic electronics promises to massively downscale the required energy per multiply-accumulate operation to a fraction of the digital electronic's energy~\cite[Fig.~1]{Prucnal17}.
Neuromorphic electronics emulate spiking neural networks (SNNs) that mimic the brain's behavior and promise energy efficiency and low-latency processing~\cite{Davies21}. 
With the rise of the first prototypes of neuromorphic hardware, e.g., Intel's Loihi~\cite{Davies21} or Heidelberg's BrainScaleS-2 system~\cite{Pehle22}, SNNs reach state-of-the-art performance in many inference tasks, like spoken number recognition~\cite{Cramer22} or hand gesture recognition~\cite{Ceolini20}, while consuming little energy~\cite{Cramer22}. 

In \cite{simeoneANN}, SNNs compress the output of a neuromorphic camera, where the compressed data is fed to an ANN for image classification. In~\cite{SimeoneSNN}, the ANN is replaced by an SNN, implementing an entire SNN-based joint source-channel coding scheme and showing the system's robustness against noisy transmission of the encoded data.
In~\cite{arnold22soft}, an SNN-based equalizer is used to mitigate the distortion of an optical channel with simple direct detection, showing that SNN-based approaches are competitive with ANN-based approaches. This SNN-based equalizer is implemented on the neuromorphic BrainScaleS-2 system in~\cite{arnold22neuro}.
However, the optical link used in~\cite{arnold22soft,arnold22neuro} mainly introduces non-linear distortions without suffering from severe intersymbol interference (ISI). 

In this paper, we introduce a novel SNN-based equalizer akin to the traditional decision feedback equalizer (DFE)~\cite[p.~707]{Proakis08} for linear channels impacted by severe ISI.
By adding a feedback path of already equalized symbols, which improves the equalizer's performance, we extend the work of \cite{arnold22soft} to enable the combat of severe ISI and propose an SNN-based equalizer and demapper with a decision feedback structure. To convert real-world data into spike signals, we introduce a novel ternary encoding.
Based on three different frequency-selective channels that furthermore experience additive white Gaussian noise (AWGN), we compare the SNN-based approach with linear equalizers (linear minimum mean square error (LMMSE) and zero-forcing (ZF)), the DFE, and ANN-based decision feedback structures.
We show that the proposed SNN-based equalizer with decision feedback can deal with severe ISI.

\newcommand{\e}{\mathrm{e}}

\section{Spiking Neural Networks}

\subsection{Spiking Neural Networks}
Like ANNs, SNNs are neural networks that interconnect neurons.
The neurons are connected via \emph{synapses}, which amplify or attenuate messages exchanged between neurons.
However, the neurons' internal dynamics as well as the kind of exchanged messages greatly differs. 
Neurons of ANNs compute a real-valued message by applying a non-linear function to the weighted sum of their inputs, which is then propagated to all downstream neurons.
Neurons of SNNs, in contrast, are integrators that leak over time. 
They have an internal state, called \emph{membrane potential} $v(t)$.
The weighted input signals are summed up and added to the membrane potential.
In parallel, the membrane potential leaks over time towards a \emph{resting potential} $v_\text{rest}$.
If the membrane potential is high enough, the neuron gets excited and generates a short output pulse, which is propagated to all downstream neurons.
The membrane potential at which the neuron gets excited and an output pulse is fired is called \emph{firing threshold} $v_\text{th}$.
After emitting the pulse, the neuron resets its membrane potential $v(t)$ to its resting potential $v_\text{rest}$~\cite{Diehl15}.
The output pulse, called \emph{spike}, is a pulse of uniform duration and amplitude, which encodes information in an ``all-or-nothing'' manner~\cite{Auge21}.
In biology, the output spike now stimulates the synapses between the actual and the downstream neuron, leading to a \emph{synaptic current} $i(t)$ of neurotransmitters that are exchanged and charge the downstream neuron's membrane potential~\cite{gerstner14}.
This way, information propagates as discrete events in an asynchronously manner through an SNN~\cite{arnold22soft}.  

\subsection{Leaky-integrate-and-fire Model}
A biologically plausible, yet easily computable neuron model ist the leaky-integrate-and-fire (LIF) model.
Its membrane potential $v(t)$ is a leaky integrator of the synaptic current $i(t)$, where both can be described by~\cite{BPTT}
\begin{align*} 
    \frac{\mathrm{d}v(t)}{\mathrm{d}t}  &=-\frac{1}{\tau_\text{m}} ((v(t)-v_\text{rest})+i(t)) \\[0.5em]
    \frac{\mathrm{d}i(t)}{\mathrm{d}t}  &=-\frac{i(t)}{\tau_\text{s}} + \sum_j w_j s_j(t) \, .
\end{align*}
The time constant $\tau_\text{m}$ describes the intensity of the leak, $v_\text{rest}$ is the resting potential.
The strength of the synapse, connecting upstream neuron $j$ to the observed neuron, is denoted by the (trainable) weight $w_j$, while $s_j(t)$ is the output spike signal generated by upstream neuron $j$.
Furthermore, $\tau_\text{s}$ is the time constant of the synapse, leading to an exponentially decaying current after each input spike~\cite{BPTT}.
Using the forward Euler method, the solution of the ordinary differential equations (ODEs) and therefore the LIF neuron's dynamics can be approximated~\cite{Cramer22}.
Assume some arbitrary initial values $v(t_0)$ and $i(t_0)$, with $t_0=0$.
The ODEs can be solved by numerical integration with fixed integration step size $\Delta t$, leading to a discrete system defined at time instants \mbox{$t=\kappa \Delta t$}, \mbox{$\kappa \in \mathbb{N}$}.
The neuron's discrete dynamics can be expressed as~\cite{norse}
\begin{align*} 
    v[\kappa+1] &= v[\kappa]\cdot \e^{-\frac{\Delta t}{\tau_\text{m}}} + i[\kappa]\cdot \e^{-\frac{\Delta t}{\tau_\text{m}}} \\[0.5em]
    i[\kappa+1] &= i[\kappa]\cdot \e^{-\frac{\Delta t}{\tau_\text{s}}} + \sum_j w_j s_j[\kappa] \, ,
\end{align*}
where $\Delta t$ can be interpreted as the system's sampling time and $v[k]$ the membrane potential at time \mbox{$t:=\kappa\Delta t$}.
If $v[\kappa]$ exceeds the neuron's threshold $v_\text{th}$, the neuron generates an output spike 
\begin{align*} 
    s_\text{out}[\kappa] = \Theta(v[\kappa]-v_\text{th}) = \begin{cases}
        1 \qquad \text{if} \quad v[\kappa]>v_\mathrm{th} \\
        0 \qquad \text{else} \quad ,
    \end{cases}
\end{align*}
where $\Theta(\cdot)$ denotes the Heaviside step function, and the membrane voltage is reset by \mbox{$v[\kappa] \leftarrow v_\text{rest}$}. 
A computational graph of an LIF neuron is given below in Fig.~\ref{fig:BPTT_graph}-(a).
Figure~\ref{fig:lif_curve} shows the behavior of an LIF neuron for a given example input spike pattern. 

Another common neuron model is the leaky-integrate (LI) model, which is often used in the last layer of an SNN trained to solve a classification task, e.g., in \cite{arnold22soft}.
It exhibits the same dynamics as the LIF neuron, however, since \mbox{$v_\text{th}=\infty$} no output spike is generated and the reset of the neuron's membrane potential is avoided, integrating the input signals with endless memory.

\begin{figure}
    \centering
    \input{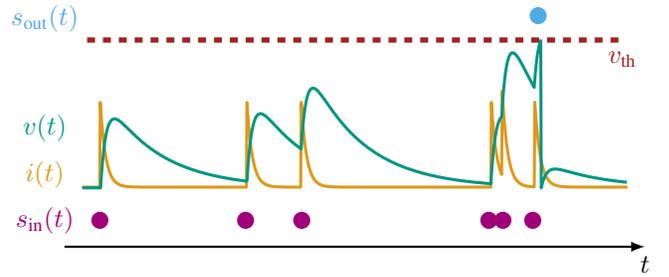}
    \caption{Example of the dynamics of an LIF-neuron. 
    The input spikes $s_\text{in}(t)$ cause the synaptic current $i(t)$, which charges the membrane potential $v(t)$ until the firing threshold $v_\text{th}$ is reached and an output spike $s_\text{out}(t)$ is fired.}
    \label{fig:lif_curve}
\end{figure}

\subsection{Recurrent Architecture}
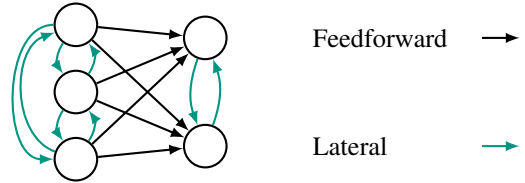
\begin{figure}
    \centering 
    \begin{tikzpicture}[>=latex,thick]
    \def\rad{0.2cm}
    \def\ydist{0.3cm}
    \def\xdist{1.3cm}

    \node[draw,circle,inner sep=\rad] (i1) {};
    \node[below=\ydist of i1, draw,circle,inner sep=\rad] (i2) {};
    \node[below=\ydist of i2, draw,circle,inner sep=\rad] (i3) {};
    
    \node[above right=\ydist and \xdist of i2, draw,circle,inner sep=\rad] (h1) {};
    \node[below right=\ydist and \xdist of i2, draw,circle,inner sep=\rad] (h2) {};

    \draw[->] (i1) -- (h1);
    \draw[->] (i1) -- (h2);
    \draw[->] (i2) -- (h1);
    \draw[->] (i2) -- (h2);
    \draw[->] (i3) -- (h1);
    \draw[->] (i3) -- (h2);

    \path[->,KITgreen] (i1) edge [bend right=35] (i2);
    \path[->,KITgreen] (i2) edge [bend right=35] (i3);
    \path[->,KITgreen] (i3) edge [bend right=35] (i2);
    \path[->,KITgreen] (i2) edge [bend right=35] (i1);
    \path[->, KITgreen] (i1) edge [bend right=90] (i3);
    \path[->, KITgreen] (i3) edge [bend left=70] (i1);

    \path[->,KITgreen] (h1) edge [bend right=20] (h2);
    \path[->,KITgreen] (h2) edge [bend right=20] (h1);

    \def\tw{1.5cm}
    \node[right=1cm of h1, text width = \tw, align=center,minimum width=\tw] (s1) {Feedforward};
    \node[right=1cm of h2, text width = \tw,minimum width=\tw] (s2) {Lateral};

    \draw[->] (s1)+(1.5cm,0cm) -- +(2cm,0cm);
    \draw[->,KITgreen] (s2)+(1.5cm,0cm) -- +(2cm,0cm);
\end{tikzpicture}
    \caption{Different connections of an SNN.}
    \label{fig:recurrent}
\end{figure}

Like ANNs, SNNs can be defined by layers.
Connecting the layers in a feedforward manner results in a feedforward SNN. 
Due to the internal dynamics of LIF neurons an SNN has implicit recurrent connections, shown in the example SNN in Fig.~\ref{fig:BPTT_graph}-(a).
Furthermore, explicit recurrent connections can be added~\cite{ZenkeNeftci}, like lateral connections, see Fig.~\ref{fig:recurrent}.
Lateral connections excite or inhibit neighboring neurons from firing~\cite{Diehl15}, feedback connections allow the neuron to excite or inhibit itself.
For classification tasks based on different datasets (Randman, MNIST, SHD, RawHD, RawSC),~\cite[Fig. 3]{ZenkeNeftci} has shown that adding explicit recurrent connections achieves a higher classification accuracy.

\subsection{Update Algorithm}
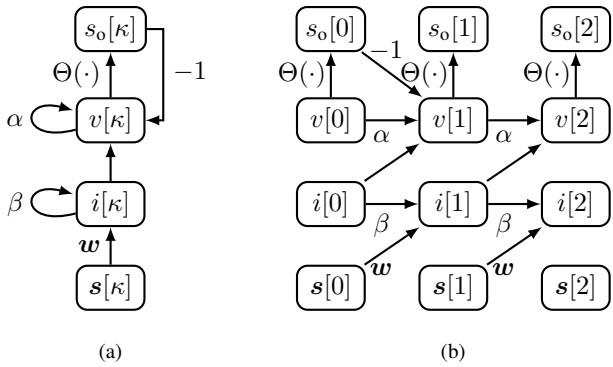
\begin{figure}
    \centering
    \begin{tikzpicture}[>=latex,thick]
    \def\dist{0.5cm}
    \def\width{0.9cm}
    \def\xdist{0.7cm}

    \node[draw, rectangle, rounded corners, minimum width=\width] (sin) {$\bm{s}[\kappa]$};
    \node[above=\dist of sin, draw, rectangle, rounded corners, minimum width=\width] (i) {$i[\kappa]$};
    \node[above=\dist of i, draw, rectangle, rounded corners, minimum width=\width] (v) {$v[\kappa]$};
    \node[above=\dist+0.1cm of v, draw, rectangle, rounded corners, minimum width=\width] (so) {$s_\text{o}[\kappa]$};

    \draw[->] (sin) -- (i) node [midway, left] (w) {$\bm{w}$};
    \draw[->] (i) -- (v);
    \draw[->] (v) -- (so) node [midway, left] (t) {$\Theta(\cdot)$};
    \path[->] (i)  edge [loop left] node {$\beta$} ();
    \path[->] (v)  edge [loop left] node {$\alpha$} ();
    \draw[->] (so) -- +(0.7cm,0cm) node [below=16pt,right] (1) {$-1$} |- (v);

    \node[right=2cm of sin, draw, rectangle, rounded corners, minimum width=\width] (sin0) {$\bm{s}[0]$};
    \node[above=\dist of sin0, draw, rectangle, rounded corners, minimum width=\width] (i0) {$i[0]$};
    \node[above=\dist of i0, draw, rectangle, rounded corners, minimum width=\width] (v0) {$v[0]$};
    \node[above=\dist+0.1cm of v0, draw, rectangle, rounded corners, minimum width=\width] (so0) {$s_\text{o}[0]$};

    \node[right=\xdist of sin0,draw, rectangle, rounded corners, minimum width=\width] (sin1) {$\bm{s}[1]$};
    \node[above=\dist of sin1, draw, rectangle, rounded corners, minimum width=\width] (i1) {$i[1]$};
    \node[above=\dist of i1, draw, rectangle, rounded corners, minimum width=\width] (v1) {$v[1]$};
    \node[above=\dist+0.1cm of v1, draw, rectangle, rounded corners, minimum width=\width] (so1) {$s_\text{o}[1]$};

    \node[right=\xdist of sin1,draw, rectangle, rounded corners, minimum width=\width] (sin2) {$\bm{s}[2]$};
    \node[above=\dist of sin2, draw, rectangle, rounded corners, minimum width=\width] (i2) {$i[2]$};
    \node[above=\dist of i2, draw, rectangle, rounded corners, minimum width=\width] (v2) {$v[2]$};
    \node[above=\dist+0.1cm of v2, draw, rectangle, rounded corners, minimum width=\width] (so2) {$s_\text{o}[2]$};

    \draw[->] (sin0) -- (i1) node [pos=0.3, below] (w0) {$\bm{w}$};
    \draw[->] (sin1) -- (i2) node [pos=0.3, below] (w1) {$\bm{w}$};

    \draw[->] (i0) -- (v1);
    \draw[->] (i1) -- (v2);

    \draw[->] (i0) -- (i1) node [pos=0.3, below] (a0) {$\beta$};
    \draw[->] (i1) -- (i2) node [pos=0.3, below] (a1) {$\beta$};

    \draw[->] (v0) -- (v1) node [pos=0.3, below] (b0) {$\alpha$};
    \draw[->] (v1) -- (v2) node [pos=0.3, below] (b1) {$\alpha$};

    \draw[->] (so0) -- (v1) node [pos=0.4, above] (10) {$-1$};

    \draw[->] (v0) -- (so0) node [midway, left=-2pt] (t0) {$\Theta(\cdot)$};
    \draw[->] (v1) -- (so1) node [midway, left=-2pt] (t1) {$\Theta(\cdot)$};
    \draw[->] (v2) -- (so2) node [midway, left=-2pt] (t2) {$\Theta(\cdot)$};

    \node[below=0.3cm of sin] () {\footnotesize (a)};
    \node[below=0.3cm of sin1] () {\footnotesize (b)};
\end{tikzpicture}
    \caption{(a) Computational graph of an LIF neuron, where $\bm{s}$ denotes the input vector of downstream neurons and $\alpha=\e^{-\frac{\Delta t}{\tau_\text{m}}}$ and $\beta=\e^{-\frac{\Delta t}{\tau_\text{s}}}$ denote the exponential decay of $v[\kappa]$ and $i[\kappa]$ respectively.
    (b) Computational graph unrolled in time, inspired by~\cite[Fig. 2]{BPTT}.}
    \label{fig:BPTT_graph}
\end{figure}

Since SNNs are time-dependent, the update rule for SNNs during training needs to take into account time~\cite{BPTT}.
Inspired by recurrent neural networks (RNNs), the computational graph of an SNN can be unrolled in time, see Fig.~\ref{fig:BPTT_graph}-(b).
Assume that \mbox{$\bm{s}_\text{o}[\kappa]$} denotes the output spikes generated at all output neurons at time $\kappa$ and \mbox{$\bm{s}_\text{t}[\kappa]$} the target spikes desired at the output.
Defining an objective function $J(\cdot)$, which compares the output spike pattern $\bm{s}_\text{o}$ with a target spike pattern $\bm{s}_\text{t}$ over time, an error at each discrete time step $\kappa$ can be calculated and backpropagated~\cite{Zenke17}.
Recall that the output \mbox{$\bm{s}_\text{o}[\kappa]$} at time $\kappa$ depends on all input spike vectors \mbox{$\bm{s}[\tilde{\kappa}],\,\tilde{\kappa}\in{0,1\ldots,\kappa-2}$}, leading to temporal interdependencies. 
Therefore, the gradient needs to be tracked over time, too.
The backpropagation through time (BPTT) algorithm solves this issue by propagating errors through the unrolled network~\cite{BPTT} akin to training of RNNs. 

However, the non-linearity \mbox{$\Theta(v-v_\text{th})$} of the LIF neuron is problematic for the application of BPTT, since its derivative is zero almost everywhere (except at $v=v_\text{th}$).
To overcome this non-differentiability of the non-linearity,~\cite{BPTT} and~\cite{Zenke17} propose to use a surrogate of the gradient during training.
The feasibility of surrogate gradients is shown in~\cite{Cramer22}, where the forward pass is executed on neuromorphic hardware, while backpropagation is done in software with the use of surrogate gradients.
The pytorch-based SNN deep learning library \texttt{norse}~\cite{norse}, which we use within this paper, is based on BPTT and surrogate gradients. 
For a more detailed description of surrogate gradients and BPTT for SNNs, the interested reader is referred to~\cite{BPTT}. 
While BPTT is a gradient based and therefore not biologically plausible, there exist several approaches of biological plausible update rules.
For further reading on the advantages and disadvantages of different update rules we refer the interested reader to~\cite{Zenke21}.

\subsection{Encoding}
\begin{figure}
    \centering
    \begin{tabular}{@{}c@{}}
        \centering
        \begin{tikzpicture}[>=latex, thick]
    \def\dist{0.3cm}
    \def\h{0.6cm}

    \node[] (yk) {$y$};
    \node[right=\dist+0.2cm of yk, draw, circle, fill=black, inner sep=1pt] (n1) {};

    \node[below right=\dist and \dist of n1, draw, rectangle,rounded corners, minimum height=\h] (sgn) {$\text{sgn}(\cdot)$};
    \node[above right=\dist and \dist of n1, draw, rectangle,rounded corners, minimum height=\h] (abs) {$\text{abs}(\cdot)$};
    \node[right=\dist of abs, draw, rectangle, rounded corners, minimum height=\h,minimum width=1.1cm] (adc) {$Q(\cdot)$};
    \node[below right = \dist-0.1cm and \dist of adc, circle, draw, inner sep=0.2mm] (n2) {\large $\times$};
    \node[right=\dist+0.2cm of n2] (enc) {$\bm{y}_\text{enc}$};

    \draw[] (yk) -- (n1);
    \draw[->] (n1) |- (abs);
    \draw[->] (n1) |- (sgn);
    \draw[->] (abs) -- (adc);
    \draw[->] (adc) -| (n2) node[pos=0.4,above] () {$\bm{\tilde{y}}_\text{enc}$};
    \draw[->] (sgn) -| (n2);
    \draw[->] (n2) -- (enc);
\end{tikzpicture} 
    \end{tabular} 
    
    \footnotesize{(a)} 
    
    \vspace{0.5cm}
    \begin{tabular}{@{}c@{}}
        \centering
        \begin{tikzpicture}[>=latex, thick]
    \def\dist{0.2cm}
    \def\disth{0.3cm}
    
    \node[] (enc1) {$y_\text{enc}^{(1)}$};
    \node[above=\dist of enc1] (enc2) {$y_\text{enc}^{(2)}$};
    \node[above=\dist of enc2] (enc3) {$y_\text{enc}^{(3)}$};
    \node[above=\dist of enc3] (encK) {$y_\text{enc}^{(4)}$};
    
    \node[right=\disth of enc1, circle, draw, inner sep=0.2cm] (i1) {};
    \node[right=\disth of enc2, circle, draw, inner sep=0.2cm] (i2) {};
    \node[right=\disth of enc3, circle, draw, inner sep=0.2cm] (i3) {};
    \node[right=\disth of encK, circle, draw, inner sep=0.2cm] (iK) {};
    \node[above=\dist of iK, text width=0.7cm] (neu) {input layer};

    \draw[->] (enc1) -- (i1);
    \draw[->] (enc2) -- (i2);
    \draw[->] (enc3) -- (i3);
    \draw[->] (encK) -- (iK);

    \node[below right=0.1cm and 0.3cm of i1] (t0) {};
    \node[right=1.8cm of t0] (t1) {};
    \draw[->] (t0) -- (t1) node[pos=0.06,below] (tstart) {$t_0$} node[pos=0.7,below] (tmax) {$t_\text{max}$};
    \draw[-] (tstart)+(0cm,0.15cm) -- ++(0cm,0.35cm);
    \draw[-] (tmax)+(0cm,0.15cm) -- ++(0cm,0.35cm);

    \node[right=2cm of t0] (t01) {};
    \node[right=1.8cm of t01] (t11) {};
    \draw[->] (t01) -- (t11) node[pos=0.06,below] (tstart1) {$t_0$} node[pos=0.7,below] (tmax1) {$t_\text{max}$} node[pos=1,below] () {$t$};
    \draw[-] (tstart1)+(0cm,0.15cm) -- ++(0cm,0.35cm);
    \draw[-] (tmax1)+(0cm,0.15cm) -- ++(0cm,0.35cm);

    \node[right=0.43cm of i1, draw, circle, KITgreen,fill=KITgreen,inner sep=2pt] () {};
    \node[right=0.43cm of i2, draw, circle, KITgreen,fill=KITgreen,inner sep=2pt] () {};
    \node[right=0.43cm of i3, draw, circle, KITgreen,fill=KITgreen,inner sep=2pt] () {};
    \node[right=0.43cm of iK, draw, circle, KITgreen,fill=KITgreen,inner sep=2pt] () {};

    \node[right=2.65cm of i1, draw, circle, KITorange,inner sep=2pt] () {};
    \node[right=2.65cm of iK, draw, circle, KITorange,inner sep=2pt] () {};

\end{tikzpicture} 
    \end{tabular} 
    
    \footnotesize{(b)}
    \caption{(a) $M$-bit ternary encoder. (b) SNN input of ternary encoded values for $M=4$.
    Filled dots denote a positive spike $y_\text{enc}^{(m)}=+1$, hollow dots a negative spike $y_\text{enc}^{(m)}=-1$.
    As an example for an input interval of $y~\in~[-2,2]$,  $y=2$ (green, left) and $y=-1.1$ (orange, right) are encoded.}
    \label{fig:tern}
\end{figure}
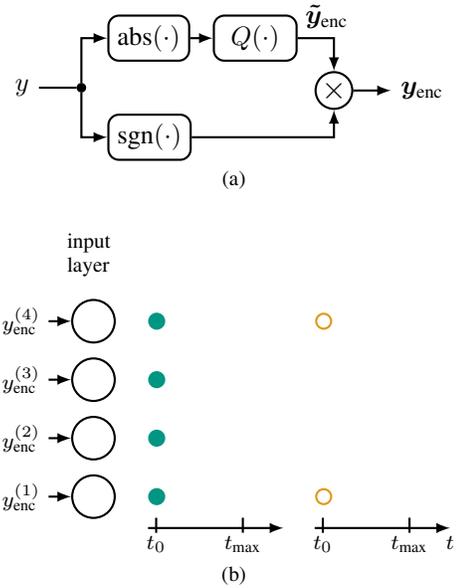
Since SNNs operate with discrete events, real world data needs to be encoded into a spike signal before processing~\cite{Auge21}.
There exist three main encoding techniques: rate, temporal or population rank coding~\cite{Petro20}. 
For rate coding, the number of spikes in a time interval (spike frequency) is proportional to the strength of the input signal that is encoded.
For temporal coding, a single spike is fired and the exact firing time of the spike with respect to a global reference, is proportional to the input~\cite{Auge21}.
To convert the input signal to a spike train, a single neuron is sufficient for rate and temporal coding.
Population rank coding uses multiple neurons, where each neuron fires a single spike and the relative time differences encode the input signal~\cite{Petro20}. 

Inspired by the success of image classification of the MNIST dataset~\cite{Diehl15}, we propose a novel encoding based on a quantizer that uses $M$ input neurons.
Each MNIST image represents a handwritten digit, ranging from 0 to 9.
In~\cite{Diehl15}, each pixel of the MNIST image represents an input neuron.
Each possible digit excites a different subgroup of input neurons, which the SNN uses for classification. 
Based on the idea that an input can be described by the triggered subgroup of input neurons, we propose a bipolar encoding described by \mbox{Fig.~\ref{fig:tern}-(a)}.
First, the value that is encoded is assumed to be within the interval $y\in[-y_\text{max},y_\text{max}]$, otherwise it is clipped.
Afterwards, the absolute value of $y$ is fed to a uniform $M$-bit quantizer \mbox{$Q(\cdot): \mathbb{R}\rightarrow \mathbb{F}^M$}, centered around $\frac{y_\text{max}}{2}$.
With the quantization resolution \mbox{$\Delta = \frac{y_\text{max}}{2^M}$} the quantized value \mbox{$\overline{y}=\left\lfloor\frac{|y|}{\Delta}+\frac12 \right\rfloor,\; \overline{y} \in \mathbb{N}$}, is converted into a binary representation using a decimal to binary conversion.
The obtained bit sequence \mbox{$\tilde{\bm{y}}_\text{enc}\in \{0,1\}^M$} is interpreted as a spike pattern. 
Depending on the sign of $y$, the spike pattern is flipped to negative values by \mbox{$\bm{y}_\text{enc}=\text{sign} (y)\cdot \tilde{\bm{y}}_\text{enc}$}, leading to a bipolar encoding, since \mbox{$\bm{y}_\text{enc}\in \{-1,0,1\}^M$}.
The $m$-th value $y_\text{enc}^{(m)}$ of $\bm{y}_\text{enc}$ is then fed to the $m$-th input neuron.
The encoding can be summarized by \mbox{$\bm{y}_\text{enc}=\text{sign} (y)\cdot Q(|y|)$}.
Figure~\ref{fig:tern}-(b) shows exemplarily how the ternary encoding is fed to the SNN.

\section{Spiking Neural Network based Decison Feedback Equalizer and Demapper}

\subsection{Simulated Communication Link}
\begin{figure}
    \centering
    \begin{tikzpicture}[>=latex,thick]
    \def\dist{0.7cm}
    
    \node[] (bin) {};
    \node[right=\dist of bin, draw, rectangle, rounded corners] (map) {\rotatebox{90}{Mapper}};
    \node[right=\dist of map, draw, rectangle, rounded corners] (can) {$h[\ell]$};
    \node[right=\dist of can, draw, circle, inner sep=0.3mm] (add) {\large $+$};
    \node[right=\dist+0.5cm of add, draw, rectangle, rounded corners, color=KITred, text height=1.6cm,align=center] (equ) {\rotatebox{90}{\shortstack{SNN-based\\DFE}}};
    \node[right=\dist+0.4cm of equ] (bout) {};
    \node[above=0.5cm of add] (noise) {$\bm{n}\sim \mathcal{N}(0,\sigma^2)$};

    \draw[->] (bin) -- (map) node[pos=0.3,above] {$\bm{b}$};
    \draw[->] (map) -- (can) node[pos=0.5,above] {$\bm{x}$};
    \draw[->] (can) -- (add);
    \draw[->] (add) -- (equ) node[pos=0.5,above] {$\bm{y}$};
    \draw[->] (equ) -- (bout) node[pos=0.7,above] {$\hat{\bm{b}}$};
    \draw[->] (noise) -- (add);
\end{tikzpicture}
    \caption{Sketch of the communication link, where $h[\ell    ]$ ist the channel's impulse response}
    \label{fig:channel}
\end{figure}
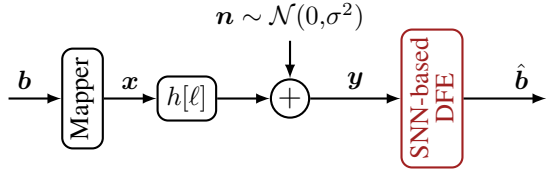

Figure~\ref{fig:channel} shows the communication link used in this paper.
A stream of random bits is Gray-mapped to a modulation alphabet $\mathcal{M}$ with $|\mathcal{M}|$ symbols. 
The transmission symbols $\bm{x}$ are transmitted over a frequency selective channel with impulse response $h[\ell]$ and AWGN is added.
Based on the received symbols $\bm{y}$, an equalizer and demapper output an estimate  $\hat{\bm{b}}$ of the transmitted bit sequence.

\subsection{Equalizer and Demapper}
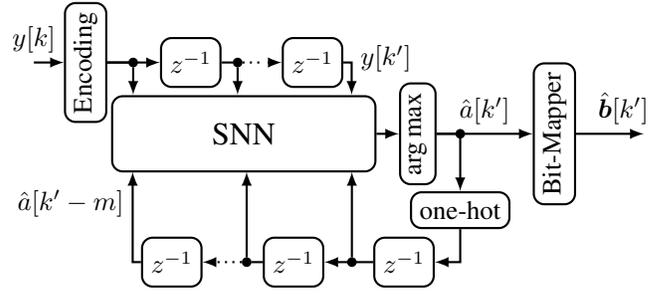
\begin{figure}
    \centering
    \begin{tikzpicture}[>=latex,thick]
    \def\zlen{0.65cm}
    \def\zdist{0.15cm}

    \node[draw,rectangle,rounded corners,minimum width=3.5cm, minimum height=1cm] (snn) {\large SNN};

    \node[above left=-0.37cm and 0.05cm of snn, draw, rectangle, rounded corners] (enc) {\rotatebox{90}{Encoding}};
    \node[left=0.4cm of enc] (y0) {};
    
    \node[draw, right=0.7cm of enc,rounded corners,minimum width=\zlen,minimum height=\zlen] (zy1) {$z^{-1}$};
    \node[draw, right=0.8cm of zy1,rounded corners,minimum width=\zlen,minimum height=\zlen] (zy2) {$z^{-1}$};
    \node[draw, right=\zdist+0.13cm of enc,circle, inner sep=1pt, fill=black] (nf1) {};
    \node[draw, right=\zdist of zy1,circle, inner sep=1pt, fill=black] (nf2) {};
    \node[below right=-0.7cm and 0.1cm of zy2] (yk) {$y[k']$};

    \node[right=0.3cm of snn, draw, rectangle, rounded corners] (argmax) {\rotatebox{90}{$\argmax$}};
    \node[right=0.25cm of argmax,draw,circle, inner sep=1pt, fill=black] (n1) {};
    \node[right=0.9 of n1, draw, rectangle, rounded corners] (demap) {\rotatebox{90}{Bit-Mapper}};
    \node[right=0.9cm of demap] (xk) {};

    \node[draw,below=0.7cm of n1,rounded corners, rectangle] (onehot) {one-hot};
    \node[draw, below left=0.1cm and -0.33cm of onehot,rounded corners,minimum width=\zlen,minimum height=\zlen] (zx0) {$z^{-1}$};
    \node[draw, left =0.6cm+0.05cm of zx0,rounded corners,minimum width=\zlen,minimum height=\zlen] (zx1) {$z^{-1}$};
    \node[draw, left =0.8cm of zx1,rounded corners,minimum width=\zlen,minimum height=\zlen] (zx2) {$z^{-1}$};
    \node[draw, left=\zdist+0.07cm of zx0, circle, inner sep=1pt, fill=black] (nb1) {};
    \node[draw, left=\zdist of zx1, circle, inner sep=1pt, fill=black] (nb2) {};
    \node[below left=0.1cm and -0.3cm of snn] (xkn) {$\hat{a}[k'-m]$};

    \def\yoff{0.43cm}
    \draw[->] (enc) -- (zy1);
    \draw[->] (enc) -- (nf1) -- +(0cm,-\yoff);
    \draw[->] (zy1) -- (nf2) -- +(0cm,-\yoff);
    \draw[dotted] (nf2) -- +(0.3cm,0cm);
    \draw[->] (nf2)+(0.4cm,0cm) -- (zy2);
    \draw[->] (zy2) -| +(0.52cm,-\yoff);
    \draw[->] (y0) -- (enc) node [pos=0.0,above] {$y[k]$};

    \def\xoff{1.2cm}
    \draw[->] (snn) -- (argmax);
    \draw[->] (argmax) -- (demap) node [midway, above] () {$\hat{a}[k']$};
    \draw[->] (demap) -- (xk) node [pos=0.7,above] {$\hat{\bm{b}}[k']$};
    \draw[->] (argmax) -| (onehot);
    \draw[->] (onehot) |- (zx0);

    \draw[->] (zx0) -- (zx1);
    \draw[->] (nb1) -- +(0cm,\xoff);
    \draw[->] (zx1) -- (nb2) -- +(0cm,\xoff);
    \draw[dotted] (nb2) -- +(-0.5cm,0cm);
    \draw[->] (nb2)+(-0.5cm,0cm) -- (zx2);
    \draw[->] (zx2) -| +(-0.52cm,\xoff);

\end{tikzpicture}
    \caption{Structure of the SNN-based DFE, where the delay block equals to the delay by one symbol.
    The received symbol $y[k]$ is delayed $n-1$ times, leading to $k'=k-n+1$, whereas $\hat{a}[k']$ is delayed $m$ times.
    Feeding in the $k$-th received sample results in the estimation of the $k'$-th transmit symbol and the corresponding bit sequence $\hat{\bm{b}}$.}
    \label{fig:snn_dfe}
\end{figure}

We replace the equalizer and demapper of Fig.~\ref{fig:channel} by an SNN, that solves both the equalization as well as the demapping task.
The structure of the SNN can be seen in Fig.~\ref{fig:snn_dfe}.
Inspired by the structure of a DFE, an $n$-tap feedforward path as well as an $m$-tap feedback path of already decided symbols are implemented.
The received samples $y[k]$ are encoded using the ternary encoding proposed in Sec.~II-E. 
Since \mbox{$y[k] \in \mathbb{C}$}, the real and imaginary parts need to be encoded in parallel.
Therefore, we encode each complex sample using $2M$ input neurons, with $M$ neurons for real and imaginary part each. 

\begin{figure}
    \centering
    \input{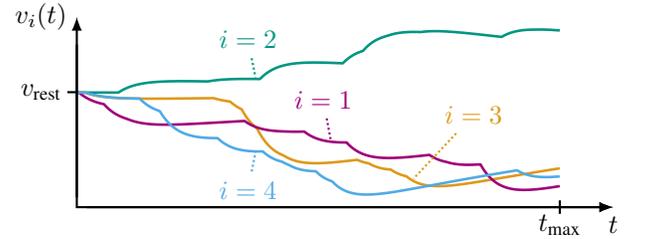}
    \vspace*{-2ex}
    \caption{LI neuron dynamics, which integrate the incoming spikes without firing, for a classification task with four classes.
    The decision would fall for the transmit symbol with index $i=2$, resulting in $\hat{a}[k]=2$.}
    \label{fig:li_curve}
    \vspace{-2ex}
\end{figure}

The SNN's output layer consists of \mbox{$N_\text{out}=|\mathcal{M}|$} LI neurons, each one representing one possible transmit symbol and allocated to an index \mbox{$i\in\{1,\ldots|\mathcal{M}|\}$}.
We simulate the SNN for $\kappa_\text{max}$ discrete time steps, resulting in a simulated time of \mbox{$t_\text{max}=\kappa_\text{max} \Delta t$}.
After $t_\text{max}$, the membrane potentials of the output neurons are read out, where the index of the neuron with the highest membrane potential indicates the index \mbox{$\hat{a}[k] \in \{1,\ldots|\mathcal{M}| \}$} of the estimated transmit symbol.
Therefore, the $k$-th transmit symbol is estimated by \mbox{$\hat{a}[k]=\argmax_i \;v_i(t_\text{max})$}, where $v_i(t)$ is the $i$-th output neuron's voltage, see Fig.~\ref{fig:li_curve}.
The estimated symbol index $\hat{a}[k]$ is demapped into the estimated bitstream $\hat{\bm{b}}$, as well as one-hot encoded and fed back to the SNN.
For one-hot encoding, each sample $\hat{a}[k]$ needs $|\mathcal{M}|$ input neurons. 
Therefore, the number of the SNN's input neurons is \mbox{$N_\text{in}=2Mn+|\mathcal{M}|m\cdot$}.

\section{Results and Discussion}
\subsection{Training Setup}
The frequency-selective channels we use are the Proakis A, Proakis B and Proakis C channels described in~\cite[p. 654]{Proakis08}.
For all three channels we compare the SNN-based equalizer with a DFE as well as two linear equalizers, the MMSE  and the ZF equalizer. 
Like in~\mbox{\cite[Fig. 9.4-4]{Proakis08}}, the linear equalizers have 31 taps. 
For better comparison, the DFE and SNN have 31 taps, too, however, the taps are split into the forward and backward path, where $m$ equals the length of the channel's impulse response and \mbox{$n=31-m$}.
We use \mbox{$M_\text{tern}=8$} input neurons for real and imaginary parts each.
For each received symbol $y$, the SNN is simulated for \mbox{$\kappa_\text{max}=10$} discrete time steps $\Delta t$, which has proven to be beneficial for simulation. The transmitted symbol is estimated as described in~Sec.~III-B.
Afterwards, the SNN is reset (i.e., all membrane potentials and synaptic currents of the SNN are reset to zero), the system's discrete time is increased by one $k \leftarrow k+1$ and the next received sample $y$ is fed to the SNN.

We use a 16-QAM constellation for the Proakis A channel and QPSK for the Proakis B and Proakis C channels.
The system parameters for each channel are summarized in Tab.~\ref{tab:channel}.
The hidden layer uses LIF neurons and the output layer LI neurons, whose paramaters are given in Tab.~\ref{tab:LIF_param}.
The hidden layer neurons are recurrently connected to all other hidden neurons. 
All SNNs are simulated using \texttt{norse} \cite{norse}, with \mbox{$v_\text{rest}=0$ and $\Delta t =10^{-3}\;\text{ms}$}.
Each SNN is trained for 10.000 epochs.
A learning rate of $10^{-3}$ is used, which is decreased by 0.08\% each epoch.
For each epoch, new training data is generated.
A burst of 200 symbols is transmitted over the channel, resulting in a batch-size of 200.
During training, we feed  the index of the correct symbol $a[k]$ back to the SNN instead of the current estimate $\hat{a}[k]$. 
During validation, the estimate $\hat{a}[k]$ is fed back.
\begin{table}[t!]
    \centering
    \caption{Architecture of the SNN-based equalizer.}
    \begin{tabular}{ccccccc}
        \toprule
         & $n$ & $m$ & $N_\text{in}$  & $N_\text{hid}$ &  $N_\text{out}$ \\
         \midrule
         Proakis A & 20 & 11 &  496 &  640 & 16\\
         Proakis B & 28 & 3 &  460 &  320 & 4\\
         Proakis C & 20 & 11 &  364 &  320 & 4\\ 
         \bottomrule \\
    \end{tabular}
    \label{tab:channel}
    \centering
    \caption{LIF and LI neuron parameters.}
    \begin{tabular}{ccc}
        \toprule
            & LIF & LI \\
         \midrule
         $\tau_\text{m}$ (ms) & 10 &  100\\
         $\tau_\text{s}$ (ms) & 5 &  1\\
         $v_\text{th}$ (V)     & 1.0 & 1000.0\\
         \bottomrule \\
    \end{tabular}
    \label{tab:LIF_param}
    \vspace*{-6ex}
\end{table}

For comparison, we train ANN-based equalizers with alike training parameters for both Proakis B and Proakis C channels.
To investigate the impact of encoding, we provide an ANN that uses ternary encoding in the forward path and one-hot encoding in the feedback path, resulting in alike architecture as the SNN. 
Furthermore, we implement an ANN whose feedforward and feedback path is without encoding, i.e., the ANN is fed with the values of $y$ and $\hat{y}$. Thus, the input layer contains $N_\text{in}=62$ neurons and an alike number of hidden and output neurons as the SNNs.
Both ANNs do not have recurrent connections and the ReLU function is used.

All networks are trained by replacing the ``arg max'' of Fig.~\ref{fig:snn_dfe} by a softmax and using the cross-entropy loss.
Furthermore, each network is trained for a fixed $E_\mathrm{b}/N_0$ value and only the synapse weights are optimized.
To evaluate the ternary encoding, we implement an SNN that uses the log-scale encoding of \cite{arnold22soft} with $M_\text{log}=10$ input neurons per encoded value for the Proakis C channel. We simulate this SNN for \mbox{$\kappa_\text{max}=30$} discrete time steps, like in~\cite{arnold22soft}.

\subsection{Results}
\begin{figure}[t!]
    \vspace{-1mm}
    \centering
    \begin{tabular}{c}
        \resizebox{1.051\columnwidth}{!}{\input{figures/res_a_channel.pgf}} \\[-0.1em]
        \small{(a) Proakis A} \\[0.2em]
        
        \resizebox{1.051\columnwidth}{!}{\includegraphics{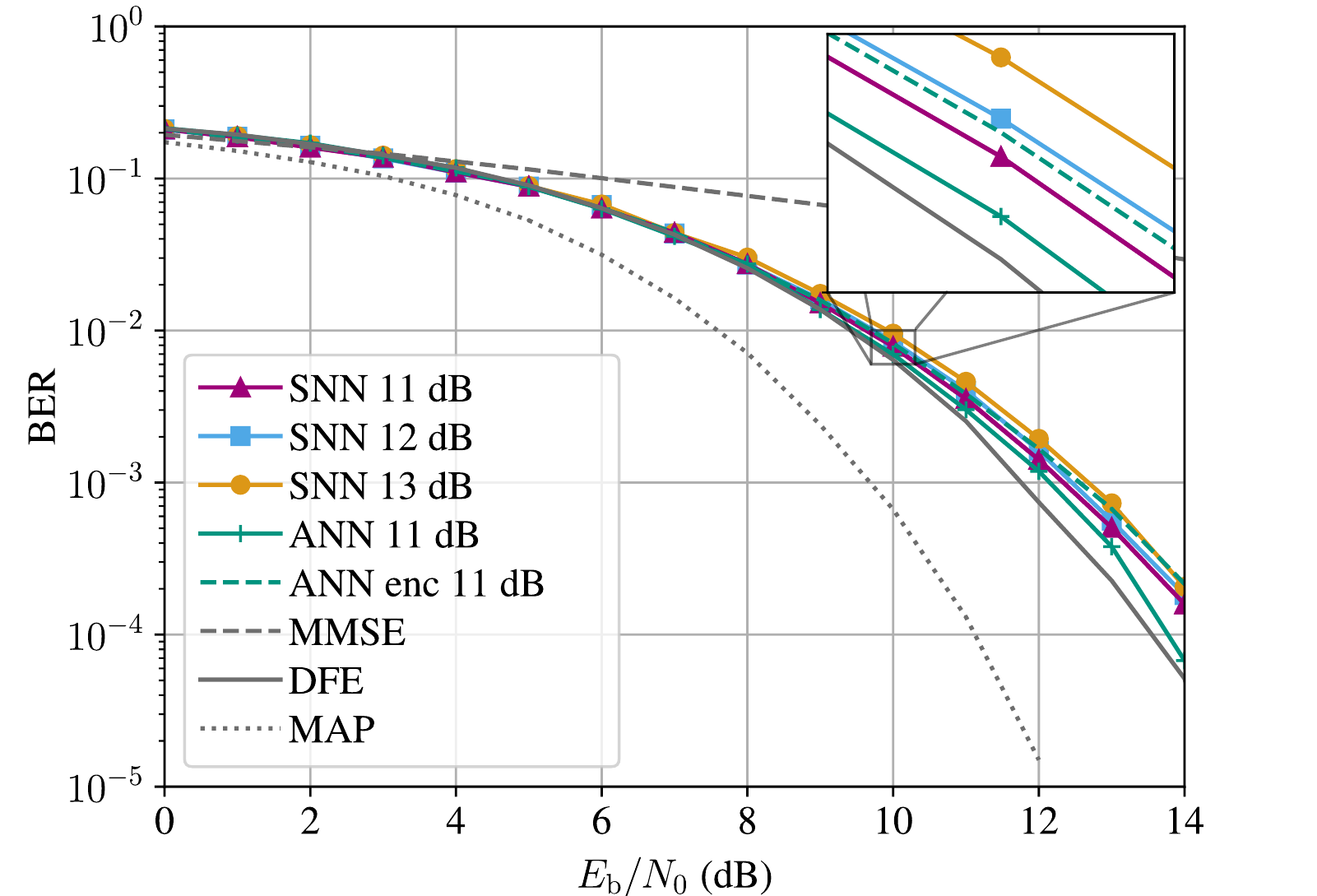}} \\[-0.1em] %
        \small{(b) Proakis B} \\[0.2em]
        
        \resizebox{1.051\columnwidth}{!}{\includegraphics{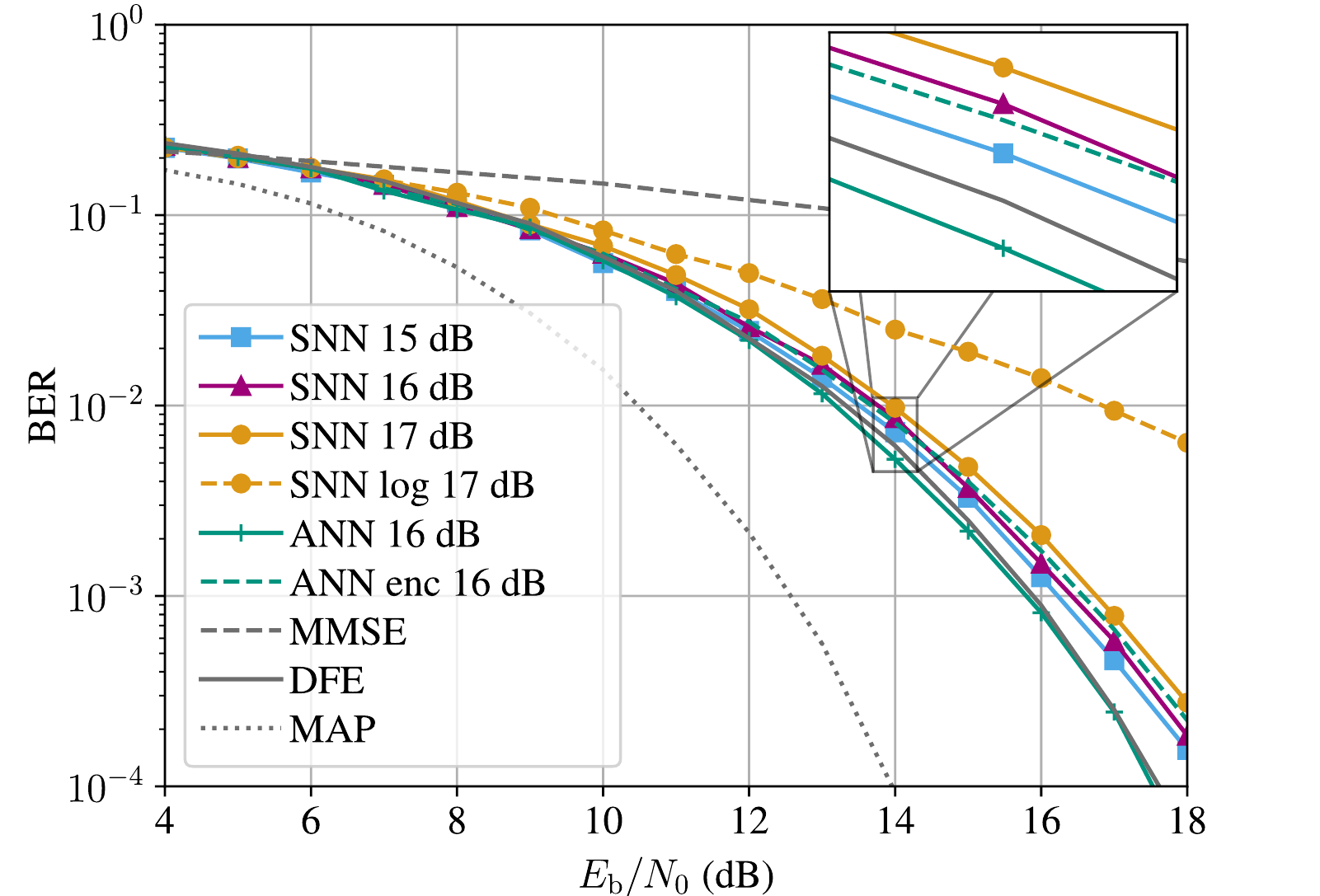}} \\[-0.1em]%
        \small{(c) Proakis C} \\[0.2em]
    \end{tabular} 
    \caption{Performance of the SNN-based equalizer for different channels.
        The $E_\mathrm{b}/N_0$ value at which  a network is trained is given behind its label, e.g., ``SNN 12\,.dB'' is trained at $E_\mathrm{b}/N_0=12\,\text{dB}$. For both Proakis B and Proakis C channels, an ANN-based equalizer with and without encoding is trained.
    ``ANN'' denotes the ANN without encoding, ``ANN enc'' the ANN with ternary encoding and ``SNN log'' the SNN with log-scale encoding.}
    \label{fig:res_dfe}
    \vspace*{-5mm}
\end{figure}
\vspace{-0.4ex}
The comparison of the benchmark equalizers as well as the ANNs with the proposed SNN-based equalizer is shown in Fig.~\ref{fig:res_dfe}.
If feasible, the MAP detector is also used as a reference.
In general, all SNNs outperform the linear equalizers.
Compared to the DFE, all SNNs have similar performance with only minor penalties.
Even when trained at a fixed $E_\mathrm{b}/N_0$, all SNNs can generalize for an arbitrary $E_\mathrm{b}/N_0$.
In Fig.~\ref{fig:res_dfe}-(b), the SNN trained at \mbox{$E_\mathrm{b}/N_0=11\,\text{dB}$} and in Fig.~\ref{fig:res_dfe}-(c), the SNN trained at \mbox{$E_\mathrm{b}/N_0=15\,\text{dB}$} outperform the SNNs trained at different $E_\mathrm{b}/N_0$.
We conclude that the SNNs can generalize for different $E_\mathrm{b}/N_0$ and that a proper choice of $E_\mathrm{b}/N_0$ for training can be better than training multiple SNNs at different $E_\mathrm{b}/N_0$.
The ANN that uses ternary encoding has worse performance than the ANN without encoding, which suggests that the encoding introduces a performance loss.
Furthermore, the SNNs outperform the ANNs that use the ternary encoding. 
The log-scale encoding, as proposed by \cite{arnold22soft}, is compared to ternary encoding in Fig.~\ref{fig:res_dfe}-(c).
With increasing $E_\mathrm{b}/N_0$, the performance gap to the SNN with ternary encoding increases.

\subsection{Discussion}
The proposed SNN-based equalizer with ternary encoding can equalize a signal distorted by a frequency-selective channel with a similar performance as a classical DFE or an ANN with a decision feedback structure.
The minor performance penalty of SNNs is mostly due to the following reasons:
First, the comparison of the ANN and the ANN with ternary encoding indicates that the encoding introduces a performance penalty.
Since the ternary encoding uses a quantizer, the encoding introduces quantization noise.
Using more encoding neurons $M$ and, therefore, more quantization steps $2^M$ could increase the resolution of the encoding, however, at the cost of a larger network architecture.
Compared to the training of the ANN, the training of the SNN is already time-consuming due to the unrolling of the SNN and the BPTT.
Therefore, more encoding neurons may minimize the encoding loss but increase the complexity. 
A more thorough investigation of the trade-off will be part of future work.
Second, due to the time constraints, we still need to fully optimize the architecture and hyperparameters.
Varying the number of hidden neurons or prolonged training may improve the SNN's performance, closing the gap to the ANN and DFE.
Furthermore, compared to ANNs, SNNs introduce new hyperparameters, e.g., the LIF parameters $\tau_\text{m}$, $\tau_\text{s}$ and $v_\text{th}$, which may be optimized during training or be subject to a more detailed hyperparameter search.  
Furthermore, the proposed equalizer applies a hard decision at the output layer.
Interpreting the membrane potential of the output neurons as soft values enables soft-decision, which may improve the error correction capability of a channel decoding applied downstream after equalization. 

Compared to the log-scale encoding of~\cite{arnold22soft}, the proposed ternary encoding appears to be more robust in this application, at least for similar training parameters, e.g., a similar number of training epochs. Finding good encoding methods is still part of ongoing research.
Furthermore, we emphasize that BPTT with surrogate gradients is a robust, yet  time-intensive update algorithm. 
More biologically plausible and faster training algorithms are objects of current research~\cite{Zenke21}, enabling the training of more complex SNNs and a more detailed hyperparameter search. 
Finally, a fair comparison of the complexity of the equalizers heavily depends on the underlying neuromorphic hardware and is beyond the scope of this work.

\section{Conclusion}
In this work we introduced an equalizer based on an SNN and inspired by the structure of the DFE.
Furthermore, we proposed ternary encoding, which is a bipolar encoding based on a quantizer.
We compared our proposed approach against linear equalizers, the DFE and ANN-based equalizers.
Our proposed approach is able to execute equalization for various frequency-selective channels, clearly outperforming linear equalizers and with similar performance than the DFE and ANN-based methods.
We furthermore showed that ternary encoding is a robust and fast encoding technique, that can outperform log-scale encoding. 
This work lays the fundamentals for future energy efficient communication receivers that use neuromorphic hardware based on spikes that promise significantly lower energy consumption than traditional receiver circuits.

\bibliographystyle{IEEEtran}
\bibliography{chapters/dfe.bib}

\end{document}